\documentclass[preprint2]{aastex7} 
\usepackage{verbatim}
\usepackage{amsmath}

\shorttitle{A $\tau - $DM relation for FRB hosts?}
\shortauthors{Mas-Ribas \& James} 

\begin{document}

\title{\large A $\tau - $DM relation for FRB hosts?}

\author[orcid=0000-0003-4584-8841]{Lluis Mas-Ribas}
\affiliation{Department of Astronomy and Astrophysics, University of California,
1156 High Street, Santa Cruz, CA 95064, USA}
\affiliation{University of California Observatories, 1156 High Street, Santa Cruz, CA 95064, USA}
\email[show]{lmr@ucsc.edu}  

\author[orcid=0000-0002-6437-6176]{Clancy W. James} 
\affiliation{International Centre for Radio Astronomy Research, Curtin University, Bentley, 6102, WA, Australia}
\email{clancy.james@curtin.edu.an}



\begin{abstract}

It has been proposed that measurements of scattering times ($\tau$) from fast 
radio bursts (FRB)  may be used to infer the FRB host dispersion measure (DM) 
and its redshift. This approach relies on the existence of a correlation between 
$\tau$ and DM within FRB hosts such as that observed for Galactic pulsars. 
We assess the measurability of a $\tau - $DM$_{\rm host}$ relation through 
simulated observations of FRBs within the ASKAP/CRAFT survey, taking into account instrumental 
effects. We show that even when the FRB hosts intrinsically follow the $\tau - $DM 
relation measured for pulsars, this correlation cannot be inferred from FRB observations; this  
limitation arises mostly from the large variance around the average cosmic DM value 
given by the Macquart relation, the variance within 
the $\tau - $DM relation itself, and observational biases against large $\tau$ 
values. We argue that theoretical 
relations have little utility as priors on redshift, e.g., for 
purposes of galaxy identification, and that the recent lack of an observed 
correlation between scattering and DM in the ASKAP/CRAFT survey is not 
unexpected, even if our understanding of a $\tau - $DM$_{\rm host}$ relation is correct.

\end{abstract}



\section{Introduction}

Fast radio bursts (FRBs) are luminous millisecond-duration radio pulses  
from extragalactic sources, currently detected up to redshift  $z\gtrsim 1$ \citep{Ryder2023}. 
Their specific origin is yet unclear, 
and they appear to arise from a variety of galactic environments, from 
dwarf to massive galaxies \citep{Gordon2023,Hewitt2024,Eftekari2024,Sharma2024,Amiri2025}. 

As opposed to pulsars in the Milky Way, the extragalactic nature 
of FRBs 
makes them unique probes of the cosmic material intersected 
by their light on its way from the source to our telescopes (in addition 
to the Milky Way and their host galaxy).  
In particular, the observable referred to as dispersion 
measure (DM) denotes the column density of free electrons 
along an FRB sightline. The total DM is a well-measured quantity owing to 
its frequency dependent $(\nu^{-2})$ delay on the arrival time of the radio signal, and 
 the fraction of DM contributed by the Milky Way can be inferred from models based 
on observations \citep{Cordes2002,Yao2017}. Furthermore, the 
average contribution from cosmic structure, i.e., the joint effect 
of the IGM and the intervening collapsed structure along the sightline, is a tracer of redshift 
and it is parameterized by the Macquart relation \citep{Macquart2020}, which depends on, and can be used to probe, cosmological parameters 
\citep[][see also the recent review by \citealt{Glowacki2024}]{Deng2014,Zhou2014,Walters2018,James2022,Wang2025}. However, 
large variance around the Macquart relation is observed, and 
the cosmic contribution is further degenerate with the dispersion 
induced by the host galaxy \citep[e.g.,][]{Ocker20222,Lee2023,Ilya2024}. 
These two effects result in the observed DM not being a precise 
predictor for the redshift of the FRB.

Recently, \citealt{Cordes2022} (see also \citealt{Ocker20222}) proposed that the DM 
contribution from FRB hosts may be inferred from the measurement of the FRB scattering timescale $\tau$  \citep[although see][for a counterexample]{Mo2025}, 
as is the case for Milky Way pulsars \citep[e.g.,][but see \citealt{He2024} for an alternate two-population scenario, and \citealt{Jing2025}]{Ramachandran1997,bhat2004,Cordes2016}\footnote{An important 
difference between pulsars and FRBs is that the trend observed for the 
first is largely governed by the fact that we observe the Milky Way from the side, edge on, while FRB hosts can have different orientations. Thus, 
a relation between scattering and DM for FRBs will not necessarily match that seen in pulsars.}.  
\cite{Cordes2022} assumed that scattering is dominated  
by the host galaxy, in agreement with recent findings 
by \citealt{Ocker2025} and \citealt{Masribas2025}, and 
proposed a theoretical $\tau - $DM model  
based on parameters 
designed to describe the interstellar medium (ISM) of 
the Milky Way from pulsar observations (i.e., their cloudlet model).  In this model, overdensities of ionised gas in cloudlets simultaneously add DM and increase FRB scattering, resulting in a correlation between total FRB DM and FRB scattering.
A precise estimate of host DM would result in better constraints for the host redshift and the cosmic contribution which, in turn, would reduce the 
uncertainties in cosmological studies \citep[see also][for constraints on host DM from galaxy observables other than scattering]{Bernales2025,Leung2025}.

However, no relation between scattering and dispersion 
measure has been observed to date for FRB hosts. In 
particular, \cite{Scott2025} have recently analyzed 
a sample of 35 high-time-resolution FRBs detected 
within the Australian Square Kilometre Array Pathfinder \citep[ASKAP;][]{Hotan2021} under 
the Commensal Real-time ASKAP Fast Transients 
\citep[CRAFT;][]{Bannister2019,Cho2020,Scott2023} survey, with 29 of these FRBs identified to a host galaxy with  redshift, and they did not find any correlation between 
these two observables (their section 4.2.2 and Figure 5). A similar 
conclusion was found by \cite{Sand2025} when assessing the morphology of 137 CHIME \citep{Chime2024} FRBs. Thus, this raises the question 
of whether a relation between host scattering and 
dispersion measure actually exists, and/or if it exists 
but its measurement is impeded due to instrumental 
effects/biases. 
 
Motivated by the aforementioned unknowns, we investigate here the observability of 
such a host $\tau - $DM relation by simulating observations of FRBs and assuming there exists  
an intrinsic relation between the two variables. 
Because instrumental effects may be important and 
a lack of correlation was just reported within the 
CRAFT survey, we consider the ASKAP telescope for 
modeling the observations. Our main conclusions, 
however, do not depend on the specific choice of instrument. 

We detail the construction of mock FRBs in  Section~\ref{sec:sims} and present our results in 
Section~\ref{sec:results}. We discuss these results and 
finally conclude in Section~\ref{sec:conclusions}.
When not explicitly stated,  DM is in units of 
pc ${\rm cm^{-3}}$, while the scattering timescale is in milliseconds 
and expressed at 1 GHz.  

\section{Mock FRBs}\label{sec:sims}

\begin{figure*}\centering
\includegraphics[width=0.75\textwidth]{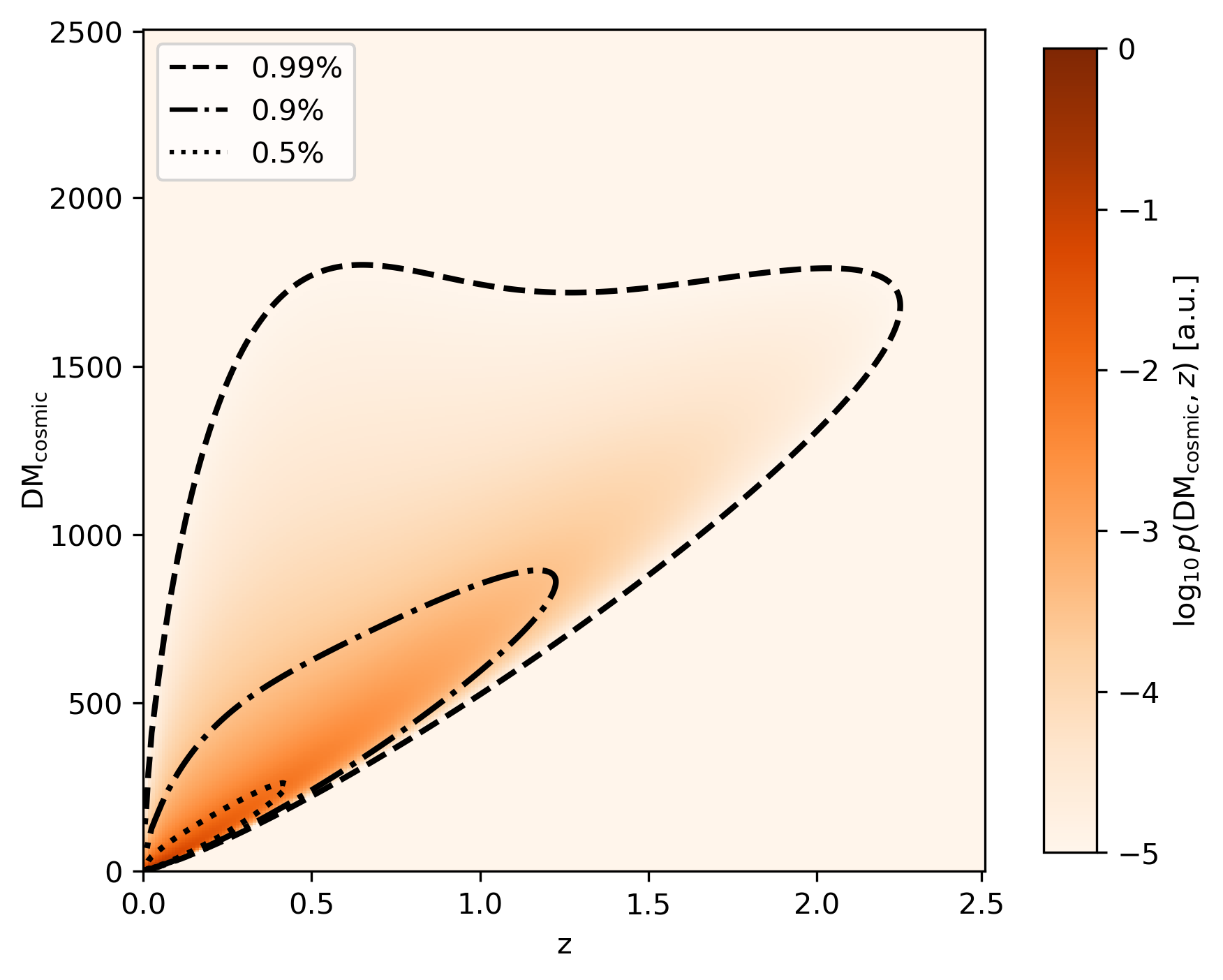}
\caption{$p(z,{\rm DM_{cosmic}})$ distribution considering the 
characteristics of the CRAFT/ICS 1.3 GHz 
instrument, and  used to create the redshift and cosmic DM values for the 25\,000 mock FRBs. The black lines denote the 50, 90 and 99 percent  contours.}
\label{fig:pzdm}
\end{figure*}

We create 25\,000 mock FRBs by first sampling pairs of DM$_{\rm cosmic}$ and FRB redshift 
$z$ from the $p(z,{\rm DM_{cosmic}})$ distribution illustrated in Figure~\ref{fig:pzdm}. 
This distribution is computed with the \texttt{zDM} code\footnote{\url{https://github.com/FRBs/zdm}} 
\citep{James2021,prochaska2023,Baptista2023} within the \texttt{FRB} library \citep{Prochaska2025},  
taking into account the characteristics of the CRAFT/ICS 1.3 GHz 
observations
\citep{Bannister2019}\footnote{CRAFT has two other frequency bands but we consider this one for simplicity. This choice does not affect our conclusions.}.  The values of the distribution arise from the best-fit model parameters of \cite{Hoffmann2025}, but with the host galaxy DM contribution, and FRB scattering, artificially set to zero.
To obtain DM$_{\rm host}$ for each of the previous FRBs, we  sample 
from a log-normal distribution centered at $\log_{10}$\,DM$_{\rm host}
=1.8$ and with $\sigma \log_{10}{\rm DM} = 0.6$ \citep{Zhang2020,Kovacs2024}. These values are 
expressed in the frame of the host, while in the frame of the observer 
they are suppressed by a redshift factor such that  
${\rm DM}_{\rm host}^{\rm obs} = {\rm DM_{\rm host}}/(1+z)$. The 
corresponding scattering times are assumed to be produced only by the host and to 
arise from the $\tau -$DM  relation for MW pulsars presented in 
\cite{Cordes2022}. We adopt this relation as our fiducial choice although 
it is well known that its shape is largely influenced by the observer 
position within the Milky Way and, therefore, FRBs may present 
a different trend. Each scattering value is drawn from a log-normal distribution centered at the value given by 
\begin{align}
    \tau_{\rm host}({\rm DM_{\rm host}},\nu) & = 5.7\times10^{-7}\, {\rm ms}\; \nu^{-\alpha} \,
       {\rm DM_{\rm host}}^{1.5} \nonumber \\
        & \times  (1 + 3.55\times 10^{-5}\, {\rm DM_{\rm host}}^3) ~, \label{eq:tauhost}
\end{align}
with $\sigma \log_{10} \tau_{\rm host}=0.76$, $\alpha=4$, and $\nu=1$ GHz.
Here above, we have multiplied the original pulsar equation by a factor of three to account for the difference between spherical and plane waves \citep{Cordes2016}\footnote{We note that this factor is only applicable under the assumption of a uniform distribution of effective lens distances between Earth and Galactic pulsars, and with an analogous geometry 
for FRBs in their host galaxies, which may not be correct in all cases. In any case, the 
presence of this factor does not impact our overall results and conclusions.}, and $\tau_{\rm host}$ and 
${\rm DM_{host}}$ are again expressed in the frame of the FRB host. 
The DM contribution from the Milky Way for each FRB is finally obtained by sampling 
a normal distribution centered at ${\rm \overline{DM}_{MW}}=80$ pc $\rm cm^{-3}$ and with $\sigma=50$ pc $\rm cm^{-3}$, 
with a minimum ${\rm DM_{MW}}= 20$ pc $\rm cm^{-3}$ (adopting moderately 
different values than these here does not impact our results).
With the above quantities, we obtain the total observed DM for each FRB as 
\begin{align}
{\rm DM}_{\rm FRB} = \frac{\rm DM_{host}}{1+z} + {\rm DM_{cosmic}} + {\rm  DM_{MW}} ~.
\end{align}

\begin{figure*}\centering
\includegraphics[width=0.5\textwidth]{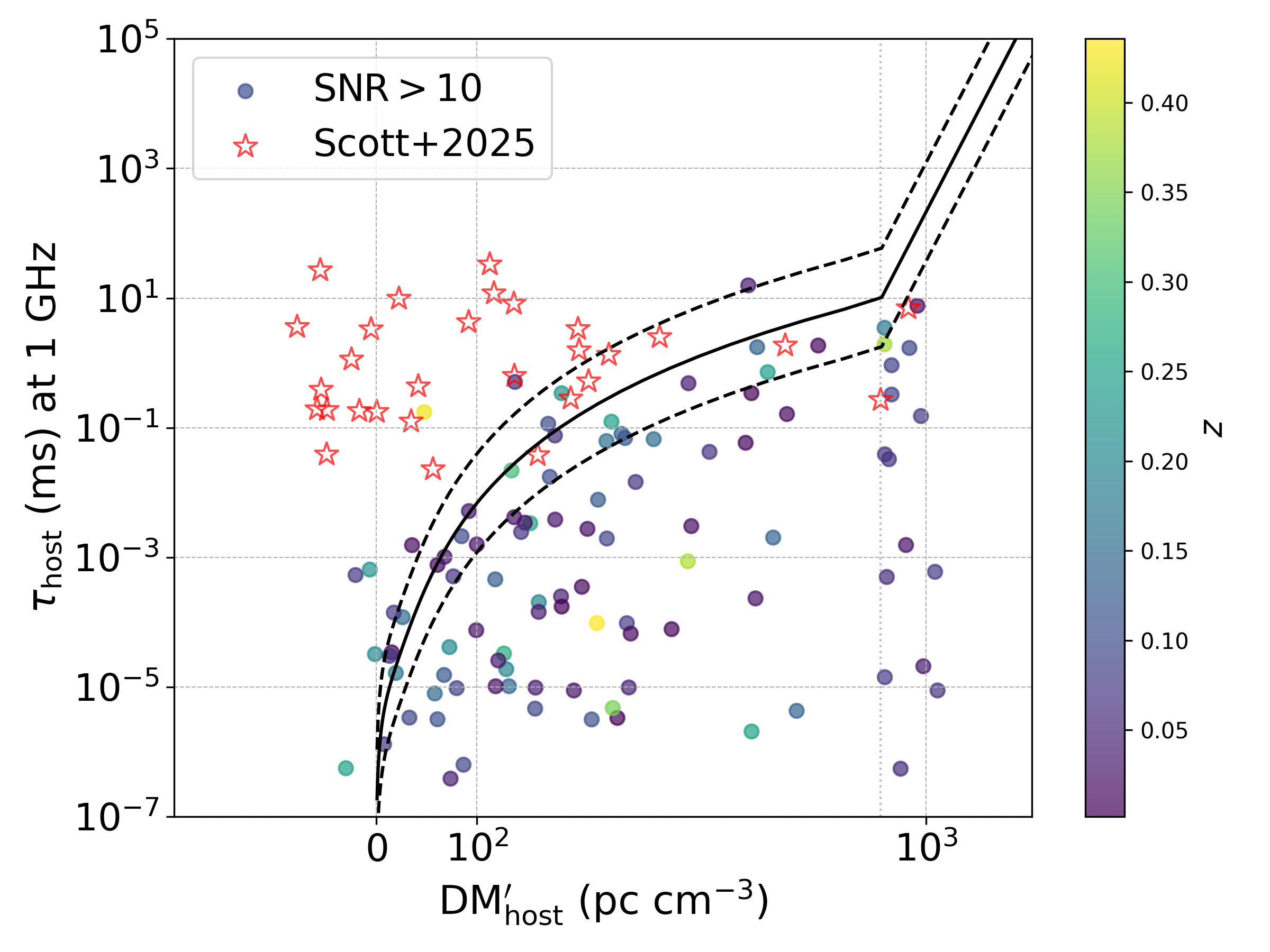}\includegraphics[width=0.5\textwidth]{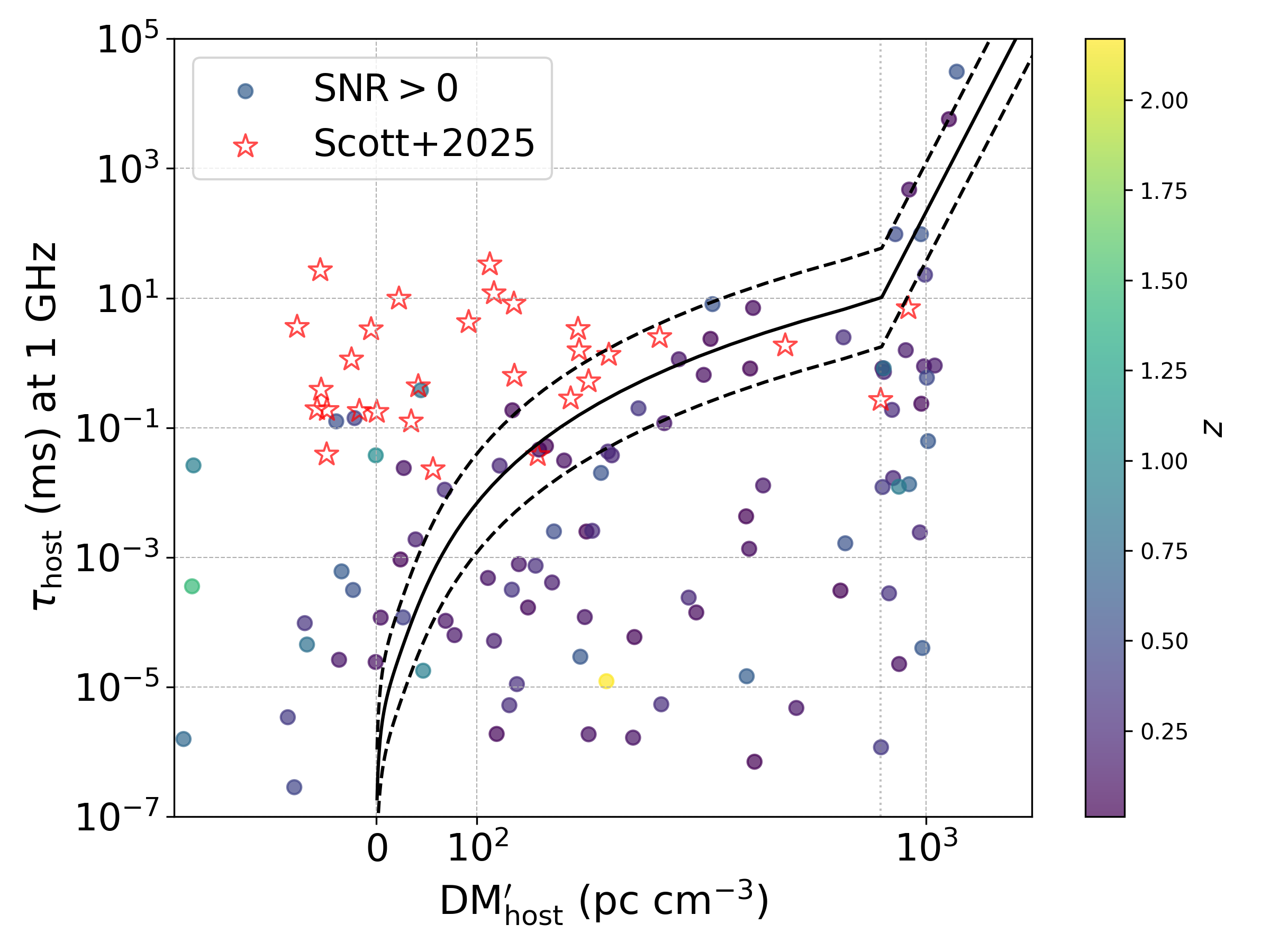}
\caption{Host scattering time against dispersion measure inferred by subtracting the Milky Way and mean cosmic components from the total DM. The figures represent a random set of 100 FRB hosts 
selected from the sample of 25\,000 mock FRBs, with an SNR threshold $\rm SNR=10$ in the left 
panel, and $\rm SNR=0$ in the right one. DM values below 500 ${\rm pc\,cm^{-3}}$ (dotted vertical line) in the horizontal axis are plotted in linear scale, and logarithmic above, for visualization. The colors indicate the redshift of the host, with different scale ranges between the two panels.
The intrinsic correlation between the scattering and the dispersion measure 
represented by the black lines (mean and 1-sigma uncertainties, respectively, from MW pulsars) 
appears largely washed out, especially for the  $\rm SNR>10$ case, where the observations are 
biased against high $\tau$ values. The stars represent the CRAFT 
data by \cite{Scott2025}, for 
comparison.}
\label{fig:taudm}
\end{figure*}

We next include the effect of observational bias on the signal-to-noise ratio (SNR) considering 
DM smearing as 
\begin{equation}
    {\rm t_{DM_{smear}}} = 8.3\, \mu s\; {\rm BW}\, \nu^{-3}\, {\rm DM_{FRB}} ~,
\end{equation}
where BW$=1$ MHz for CRAFT/ICS. The true SNR is then 
\begin{equation}
    \left[{\frac{\rm SNR}{{\rm SNR}_{\rm ins}}}\right]^4 = \frac{t_s^2 + {\rm {t_{DM_{smear}}}}^2}{t_s^2 + {\rm {t_{DM_{smear}}}}^2 
    + \left[\tau_{\rm host}\, (1+z)^{-3} \right]^2} ~, \label{eq:degrade} 
\end{equation}
where ${\rm SNR}_{\rm ins}$ is given by the instrument and $t_s=1.182$ ms is the 
sampling time \citep{Hoffmann2025}.  The factor of $(1+z)^{-3}$ multiplying $\tau_{\rm host}$ accounts for the frequency-dependence in Eq.~\ref{eq:tauhost} of $\nu^{-4}$, and the Doppler broadening factor of $1+z$. We note that this scaling is approximate only \citep[see ][ for a discussion]{2024MNRAS.528.1583H}.
We require this procedure because the \texttt{zDM} code has already accounted for the time-smearing in the numerator on the right-hand-side of Eq.~\ref{eq:degrade}, but not the additional smearing due to scattering. FRBs with resulting SNR below a given threshold are assumed to be undetectable.

With the dispersion measure created for all mock FRBs, we can 
now obtain the value of ${\rm DM_{host}^{\prime}}$ that would be inferred from observations. 
The contributions of the Milky Way and the cosmic material are not 
known with precision, so we estimate their averages.  The cosmic 
contribution to DM is estimated by adopting the average value of the  
Macquart relation \citep{Macquart2020} at the redshift of the FRBs (${\rm \overline{DM}_{cosmic}(z)}$) as is common procedure in the literature, assuming that 
$z$ is known. For the Milky Way value, we again adopt ${\rm  \overline{DM}_{MW}}=80$ 
pc ${\rm cm^{-3}}$ for all FRBs. Quantitatively, the inferred host 
DM equates 
\begin{equation}\label{eq:dmhostprime}
    \frac{\rm DM_{host}^{\prime}}{1+z} = {\rm DM_{FRB} - \overline{DM}_{cosmic}(z) - \overline{DM}_{MW}} ~.
\end{equation}

\section{Results} \label{sec:results}

Figure~\ref{fig:taudm} shows 100 random FRBs selected from the sample of 25\,000 mock FRBs, 
with an SNR threshold $\rm SNR=10$ in the left panel, and $\rm SNR=0$ in the right one. DM values below 500 ${\rm pc \, cm^{-3}}$ (marked with a vertical dotted line) are plotted in linear space and logarithmic above, for visualization.
To convert from the observed to the host frame, we assumed that the FRB 
redshift is known. This redshift is denoted by the color code; because   ${\rm DM_{host}}$ depends on ${\rm DM_{cosmic}}$ via Equation~\ref{eq:dmhostprime}, and the latter depends on redshift, one may envisage a relation between the redshift and ${\rm DM_{host}}$, although this is not apparent in our plots (we 
revisit this point in Section~\ref{sec:dependence}). Overall, the intrinsic correlation between the scattering and the dispersion measure 
represented by the black lines (mean and 1-sigma uncertainties, respectively, from MW pulsars) 
appears largely washed out, especially for the  $\rm SNR>10$ case, where the observations are 
biased against high scattering time values. 
The majority of data points appear below the pulsar relation. This is 
largely driven by the asymmetry toward large DM values around 
the Macquart relation in the $p(z,{\rm DM_{cosmic}})$ distribution in Figure~\ref{fig:pzdm}. In other words, assuming cosmic DM values 
from the Macquart relation often results in overestimated DM values 
for the host as expected from the pulsar relation. Furthermore, these data differ  
significantly from the CRAFT \citep{Scott2025} observations denoted 
by the stars.

\begin{figure*}\centering
\includegraphics[width=0.5\textwidth]{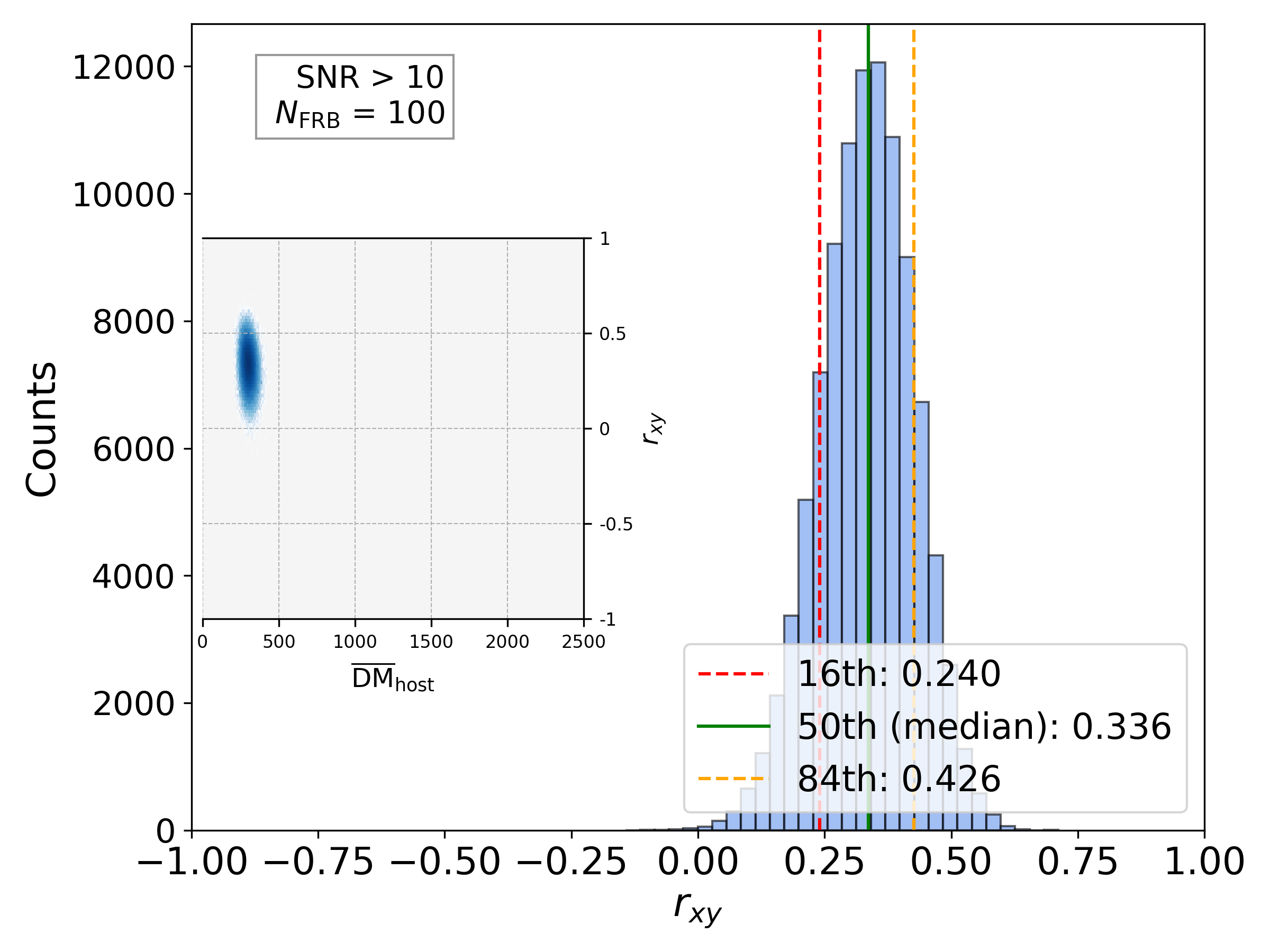}\includegraphics[width=0.5\textwidth]{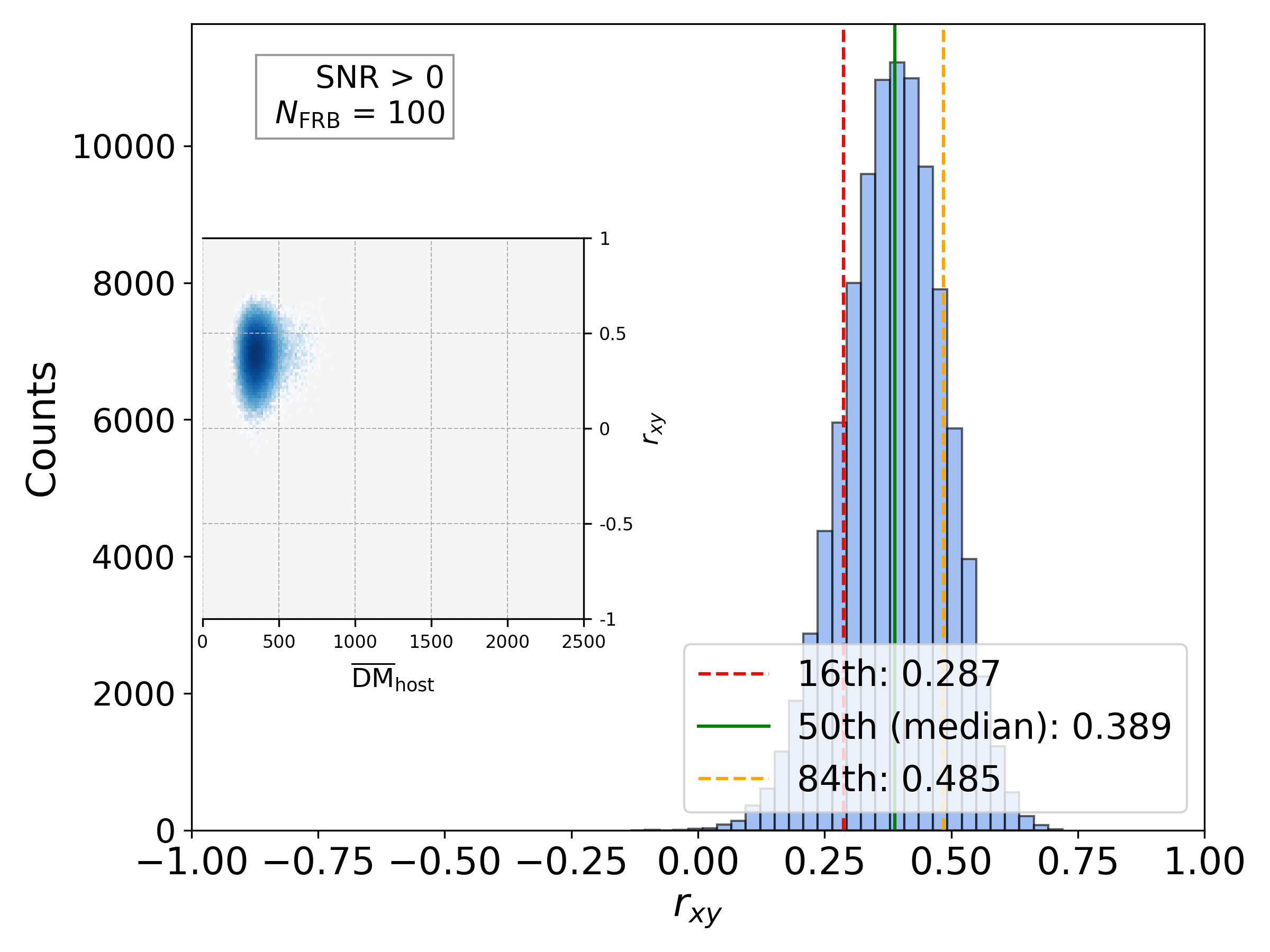}
\caption{Distributions of correlation coefficients between 
$\tau$ and ${\rm DM}$ for $10^5$ 
realizations of one hundred FRB host observations 
drawn from our mock dataset. The left panel shows the cases with 
an SNR threshold of 10, while the right panel illustrates those with no threshold. Most realizations (84th percentile) present values below 0.5, 
indicating a weak correlation between scattering and DM. Measurements containing large average host DM values may show an apparent correlation (inset figures), but this occurs a small number of times.}
\label{fig:corr}
\end{figure*}

Figure~\ref{fig:corr} illustrates the distributions of Spearman  rank-order correlation coefficients between 
$\log_{10}\tau$ and ${\log_{10}\rm DM}$ for $10^5$ 
realizations\footnote{For this calculation we use only data where 
DM is positive. Our results do not change when considering instead linear 
values (positive and negative) for the two observables in the 
correlation calculations.} of 100 FRB host observations 
drawn from our mock dataset. The left panel denotes the case with an 
SNR threshold of 10, while there is no threshold in the right one. This figure shows that when there is a statistically 
significant number of measurements (one hundred), the correlation coefficient in most realizations appears below $\sim 0.5$ (84th percentile), indicating 
a weak correlation between the two variables. Furthermore, the correlation 
is, on average, even weaker when considering an SNR threshold. This is due 
to the suppression of large DM and scattering values which, otherwise, may 
sometimes boost the correlation to higher values. This is visible in the 
inset figure of the right panel, which shows the mean host dispersion measure 
in each set of measurements and the corresponding correlation coefficient.  Overall, a strong correlation between scattering and DM cannot be 
inferred for most realizations. A smaller sample size may show a broader 
distribution of values, but those results may lack statistical significance, the exact value depending on the specific number and correlation 
strength \citep{Bonett2000}.

In addition to the mock FRBs with an intrinsic correlation mimicking the one for MW pulsars, 
we have created another dataset of mocks based on the model by \cite{Cordes2022} for comparison. 
In brief, this model makes use of a parameterization of the physical conditions in the MW ISM to 
obtain a relation between scattering and dispersion measure. Quantitatively, this expression 
equates 
\begin{align}
    \tau({\rm DM},\nu) = 48.03\, \mu s\, \nu^{-4} A_{\tau} \widetilde{F} G\, {\rm DM}^2~,
\end{align}
where $A_{\tau} \approx 1$, $\widetilde{F}$ parametrizes  the density fluctuations 
driven by ISM turbulence, and $G$ is the geometric term accounting for the relative positions 
of the FRB, the scattering material and the observer \citep[see][for 
 detailed descriptions of these parameters]{Ocker20222}. Following these authors, we 
express the product $\widetilde{F}G$ as  a uniform 
distribution $X\sim U(0.01, 10)$ in units of $({\rm pc^2\,km})^{-1/3}$. Because we sample from this distribution spanning two decades  
when computing the scattering given DM values, this range is expected to already wash out 
the relation between the two observables of interest. As for the pulsar case, 
the results for this relation are shown in Figure~\ref{fig:taudmc22} and 
in the Appendix in Figure~\ref{fig:corrc22}, where we find similar conclusions as before. 
In this case, however, the simulated data 
does match the parameter space covered by the 
CRAFT measurements better than with the pulsar 
relation. 

\begin{figure*}\centering
\includegraphics[width=0.5\textwidth]{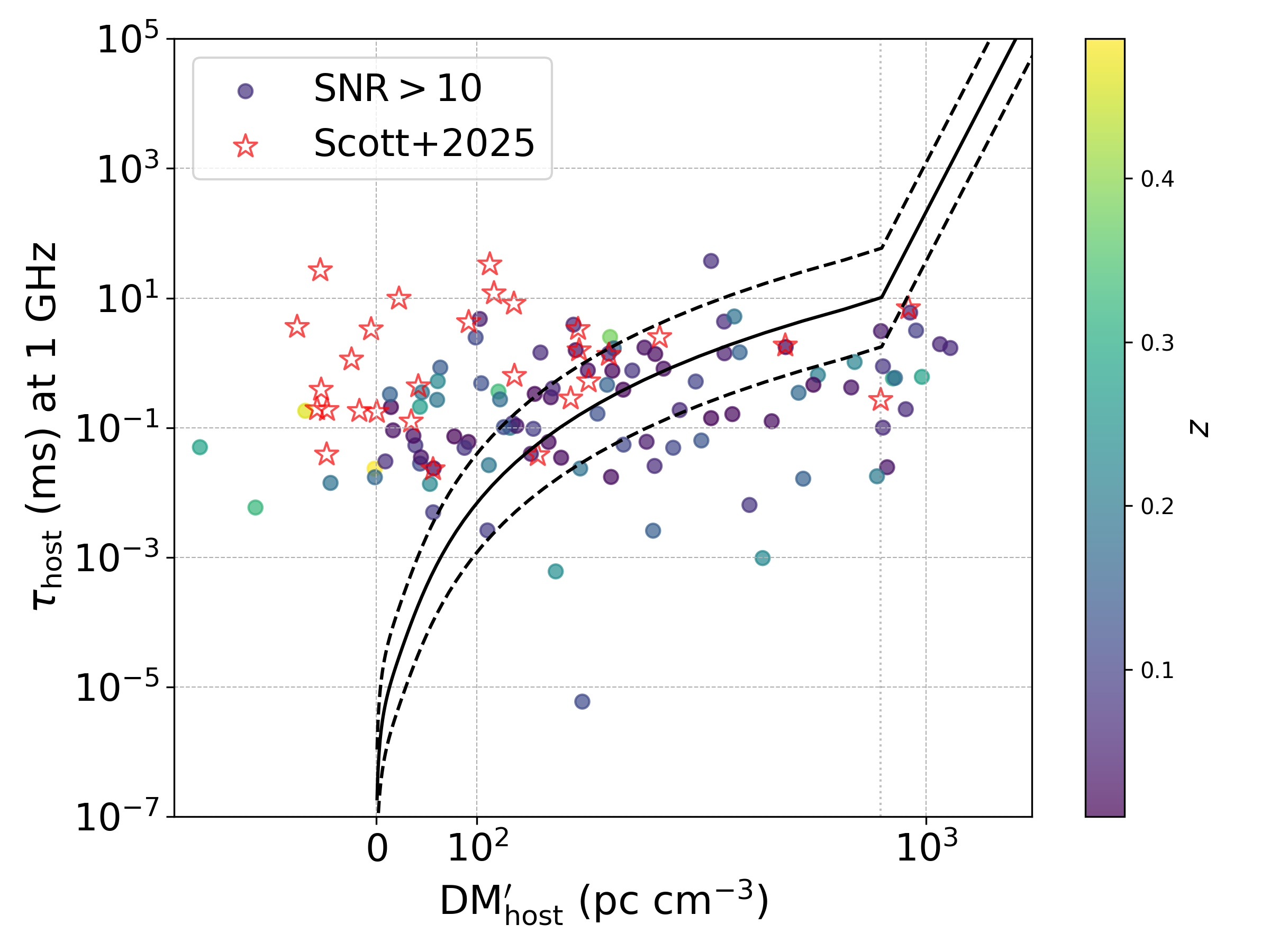}\includegraphics[width=0.5\textwidth]{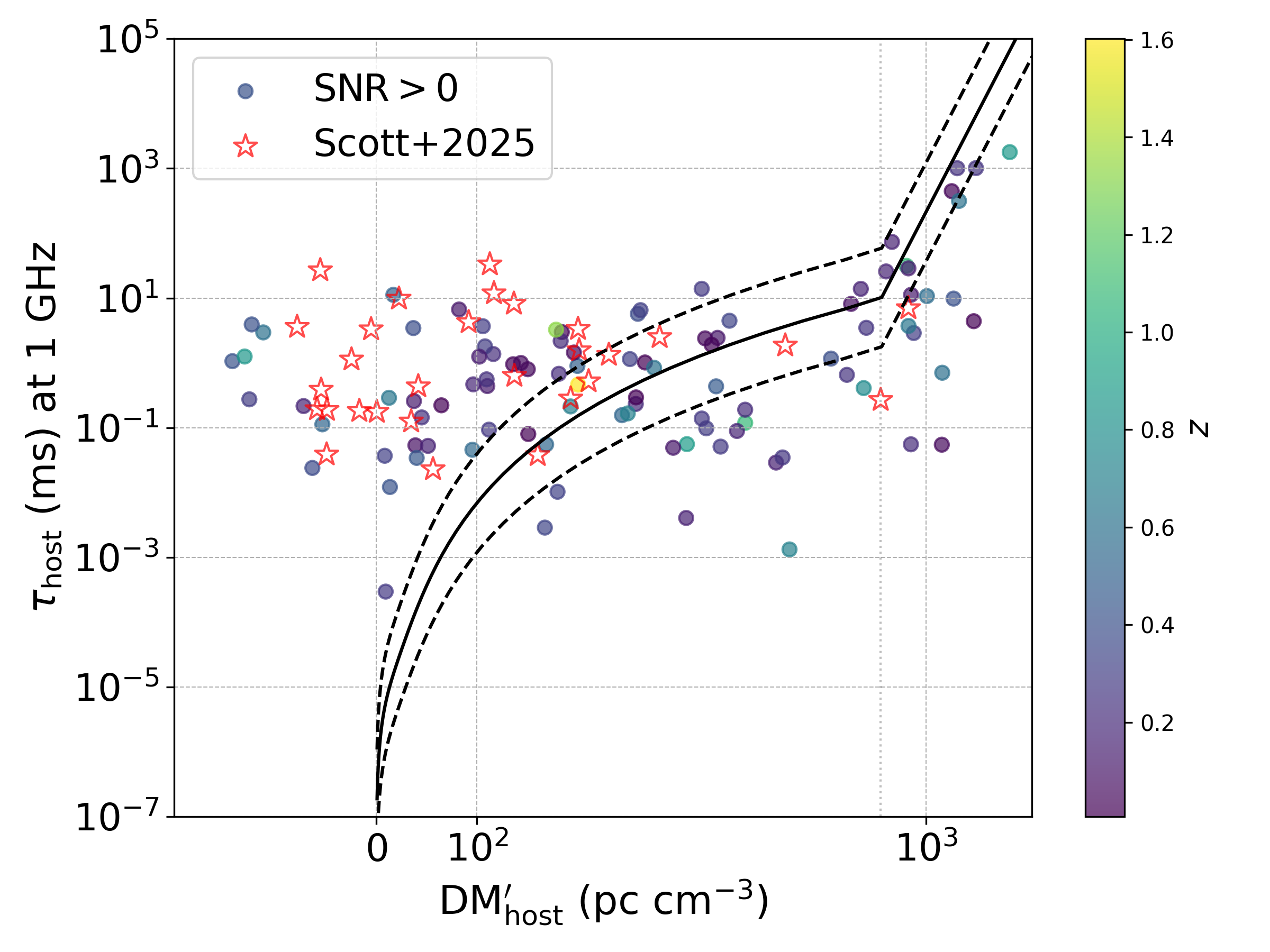}
\caption{Same as Figure~\ref{fig:taudm} but considering the $\tau - {\rm DM}^2$ relation arising 
from the cloudlet model by \cite{Cordes2022} and a uniform distribution for their geometric 
and turbulent parameters.}
\label{fig:taudmc22}
\end{figure*}

\subsection{Observability dependence on z and ${\rm DM_{FRB}}$} \label{sec:dependence}

\begin{figure*}\centering
\includegraphics[width=0.5\textwidth]{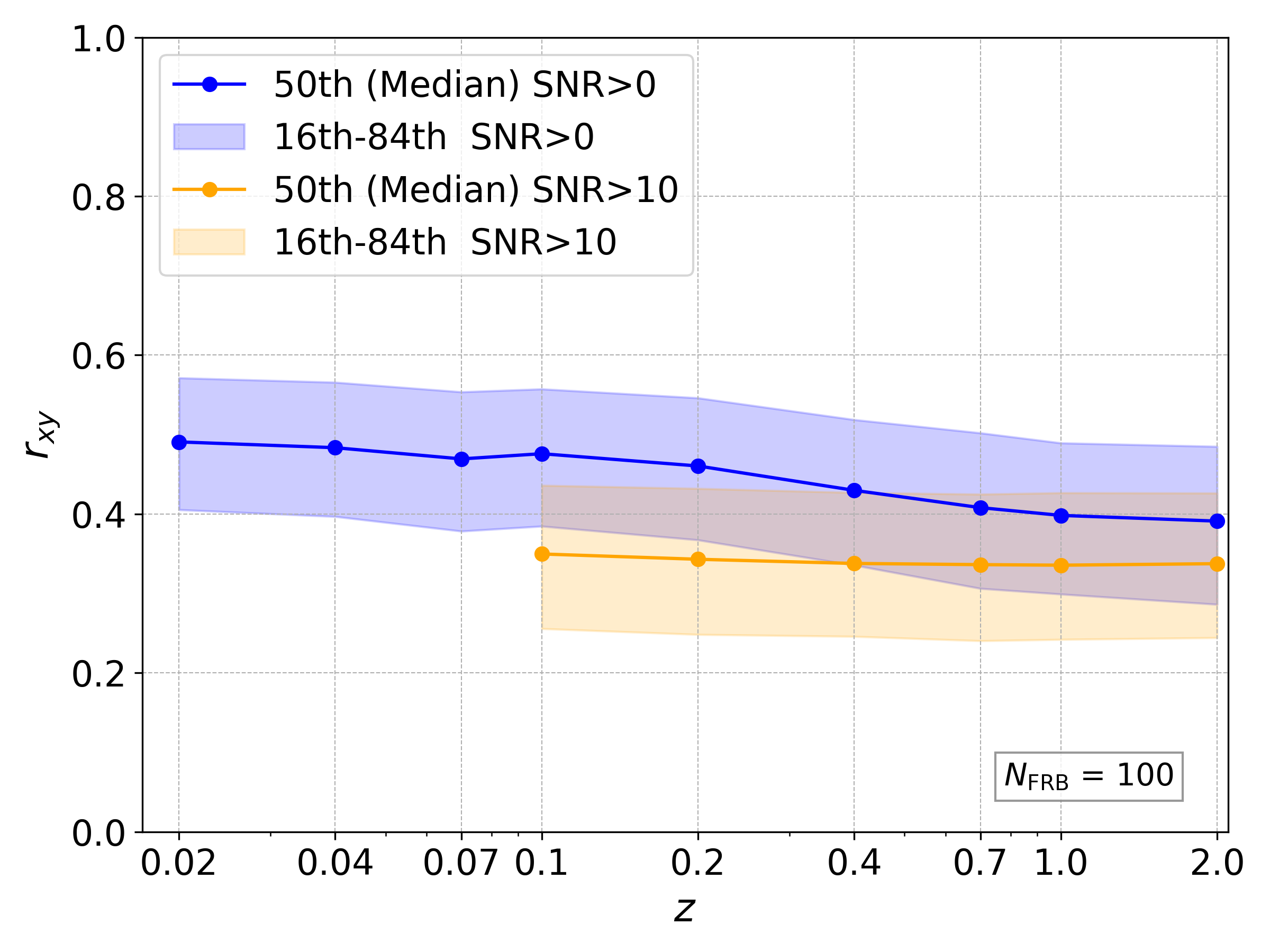}\includegraphics[width=0.5\textwidth]{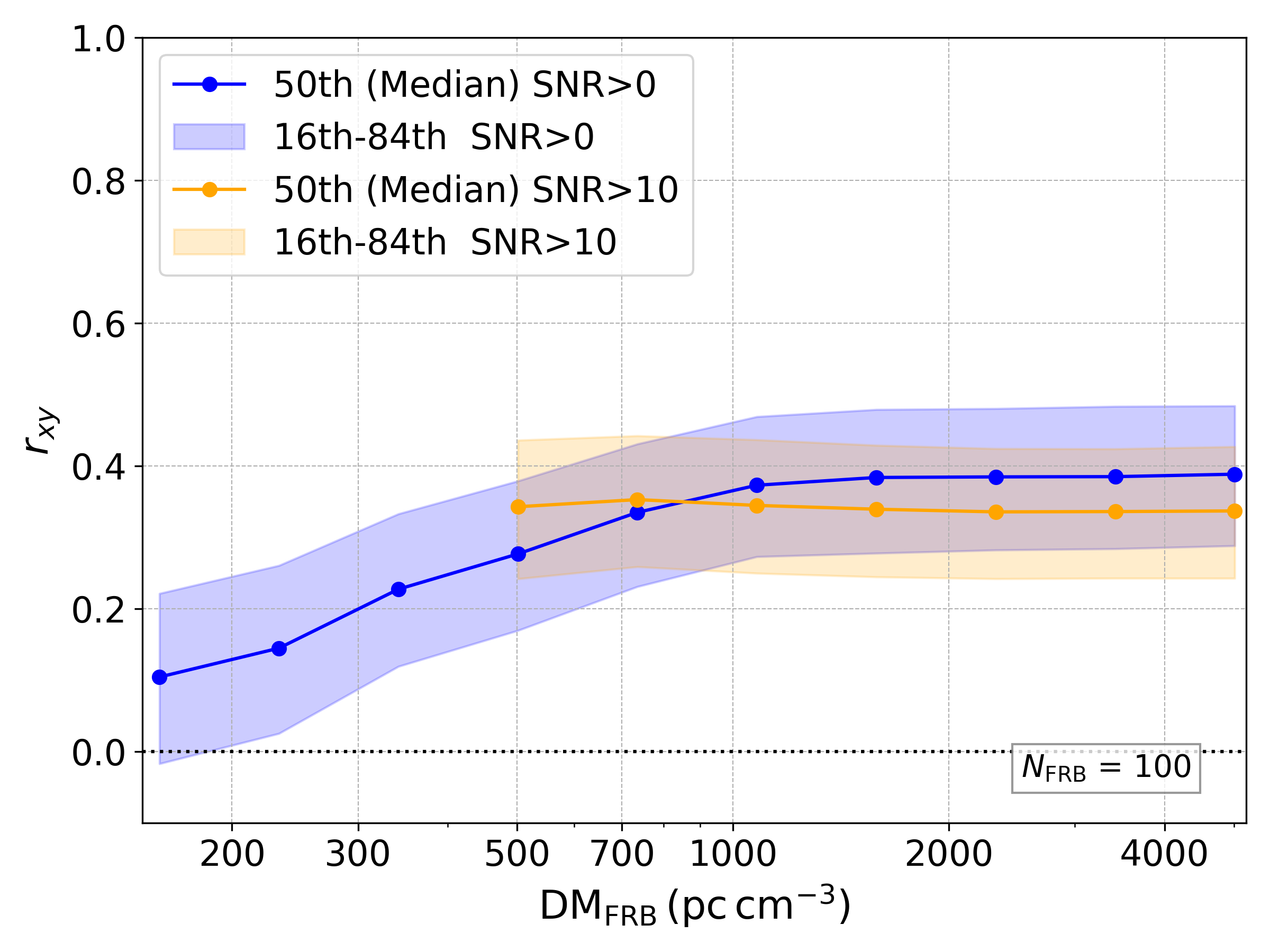}
\caption{Correlation 
coefficient with respect to FRB redshift (left) and DM (right) for the two SNR cuts   
(SNR$>10$ in orange and SNR$>0$ in blue). The SNR$>10$ cases 
do not extend below $z=0.1$ because of the small number of data 
points in those subsamples. The region $z>2$ effectively shows the same 
results as $z=2$. In no case the distributions reach correlation values 
above 0.6, indicating that no strong correlation can be inferred from the 
observations.}
\label{fig:zdm}
\end{figure*}

In the previous section we found that the observability of a 
$\tau-$DM$_{\rm host}$ relation may be more likely when the data 
contains large DM$_{\rm host}$ (and corresponding $\tau$) values. 
We examine here  the dependence of such a measurability on two other observables connected to DM$_{\rm host}$ to further explore this 
finding: a) Because the variance of DM$_{\rm cosmic}$ around the 
Macquart relation 
is a major contributor for washing out the internal $\tau-$DM$_{\rm host}$ relation, observations at (very) low redshift ($z\lesssim 0.1$),  
where the DM$_{\rm cosmic}$ variance is low (Figure~\ref{fig:pzdm}),  
may aid at detecting the relation 
of interest; b) Because the source redshift  in a) is typically 
not known, one may, a priori, consider a selection based on total 
${\rm DM_{FRB}}$ to obtain a similar effect. Specifically, the 
quantity of interest is the extragalactic DM and not the total DM, but 
in our formalism the two are just different by a constant.

The left panel in Figure~\ref{fig:zdm} illustrates the dependence of the correlation 
coefficient  on FRB redshift for the same SNR thresholds  
and number of FRBs per observation as before. The SNR$>10$ cases 
do not extend below $z=0.1$  because of the small number of data 
points in those subsamples. Overall, a slight increase of about 0.15 
in correlation coefficient at low z is visible for the SNR$>0$ case, 
but the  values ($r_{\rm XY}<0.6$) are still far from a robust confirmation of a correlation. The right panel 
displays the correlation coefficients 
with respect to  ${\rm DM_{FRB}}$. The SNR$>0$ cases, which 
contain a large enough number of data points in each subset, show a 
reduction of $\sim 0.2 - 0.3$ in the correlation values from the 
largest to the smallest ${\rm DM_{FRB}}$. Although lower 
${\rm DM_{FRB}}$ values generally correspond to lower redshifts, 
which as just shown above may slightly boost the detectability, 
these also imply that the DM$_{\rm host}$ values 
are (on average)  reduced. This reduction of DM$_{\rm host}$ has a stronger effect than that of redshift 
and drives the decline of the correlation towards low ${\rm DM_{FRB}}$. 

Figure~\ref{fig:zdmc22} corresponds to the two 
aforementioned calculations for  the Cordes et al. model, which yields the 
same conclusions just stated for the pulsar relation.

\section{Conclusions} \label{sec:conclusions}

In this paper we have explored the observability of a $\tau - $DM  relation for FRB hosts 
which, if existent,  could aid at determining the host redshift and reducing uncertainty 
in cosmological studies that depend on the cosmic DM contribution. We have created a large 
mock FRB host dataset by assuming that there exists an intrinsic $\tau - $DM$_{\rm host}$ relation 
as that observed for pulsars, as well as one that follows the modeling by \cite{Cordes2022}. 
To account for instrumental systematics, we have considered observations of such FRBs within 
the ASKAP/CRAFT survey. Our results can be summarized as follows:

\begin{itemize}
    \item[1.] Even when a tight relationship between scattering and dispersion measure from the host exists, 
    this is generally not visible/measurable from FRB observations, due 
    to fluctuations in DM from the Milky Way, intervening halos and the IGM. 

    \item[2.] Only observations including a wide range in DM$_{\rm host}$  that spans to large values may 
    enable the measurement of a correlation between the two variables resembling that 
    observed for pulsars. Observations, however, 
    are biased against high scattering (and in turn DM) values, thus hindering a possible 
    actual measurement. 
\end{itemize}

The lack of an observable relationship between scattering and dispersion measure in FRBs 
cautions against the use of relations based on Milky Way parameters. Conversely, this also means that the observed lack of a correlation, as found by \cite{Scott2025}, does not argue against the existence of an intrinsic $\tau -$DM$_{\rm host}$ relation, such as that proposed by \cite{Cordes2022}.

\begin{acknowledgments}
We are grateful to Mawson Sammons and Robert Main for discussions on scattering, and  Marcin Glowacki, Nicol\'as Tejos, Wen-fai Fong, Xavier  Prochaska, Alexa Gordon, Ilya Khrykin as well as the members of the CRAFT and 
F4 collaborations for their ideas, comments and suggestions on this work. 
L.M.R, as a member of the
Fast and Fortunate for FRB Follow-up (F4) team, acknowledges 
support from NSF grants AST-1911140, AST-1910471, and
AST-2206490. C.W.J  acknowledges support through Australian
Research Council (ARC) Discovery Project (DP) DP210102103.
\end{acknowledgments}

\appendix


\begin{figure*}\centering
\includegraphics[width=0.5\textwidth]{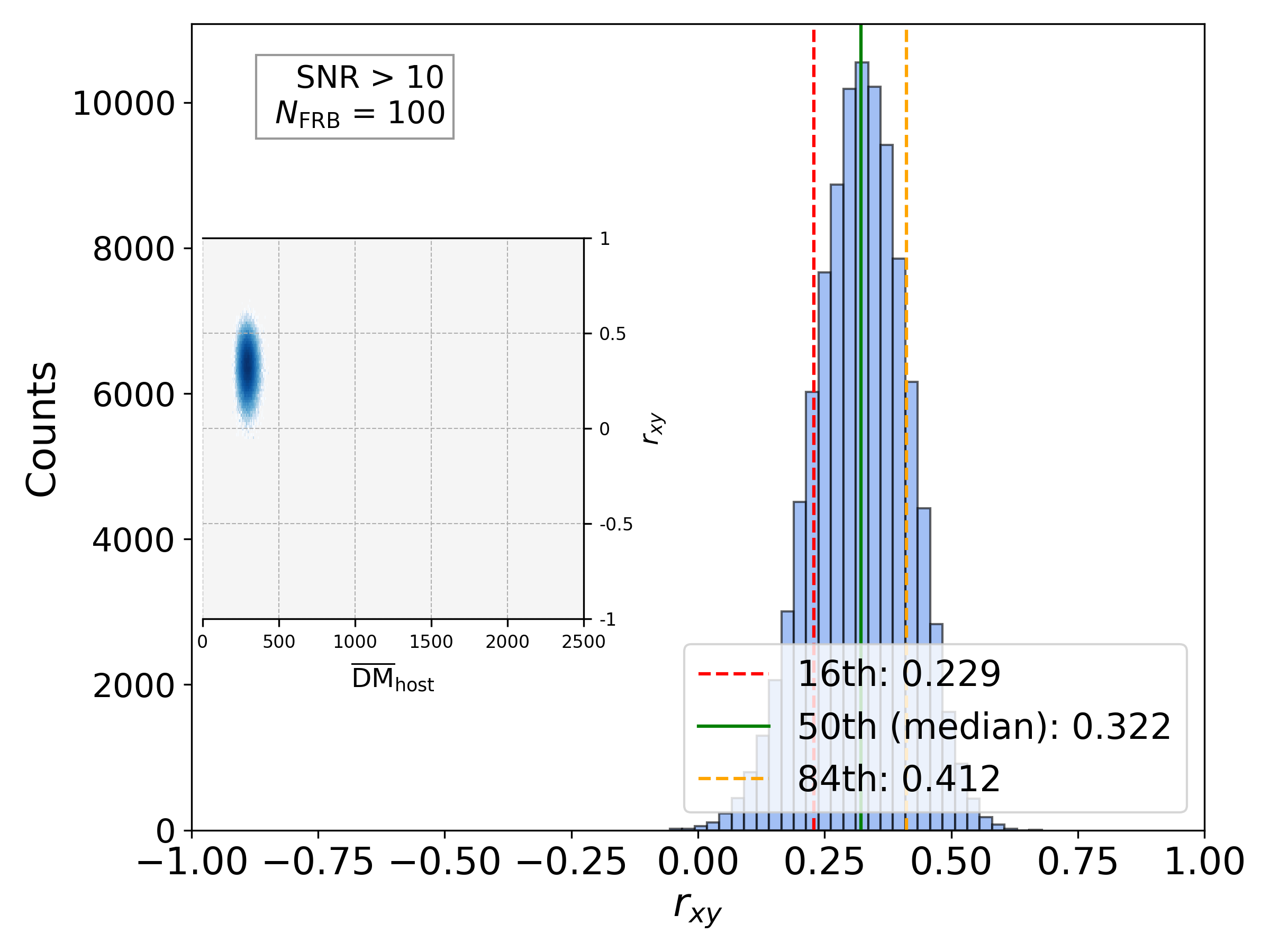}\includegraphics[width=0.5\textwidth]{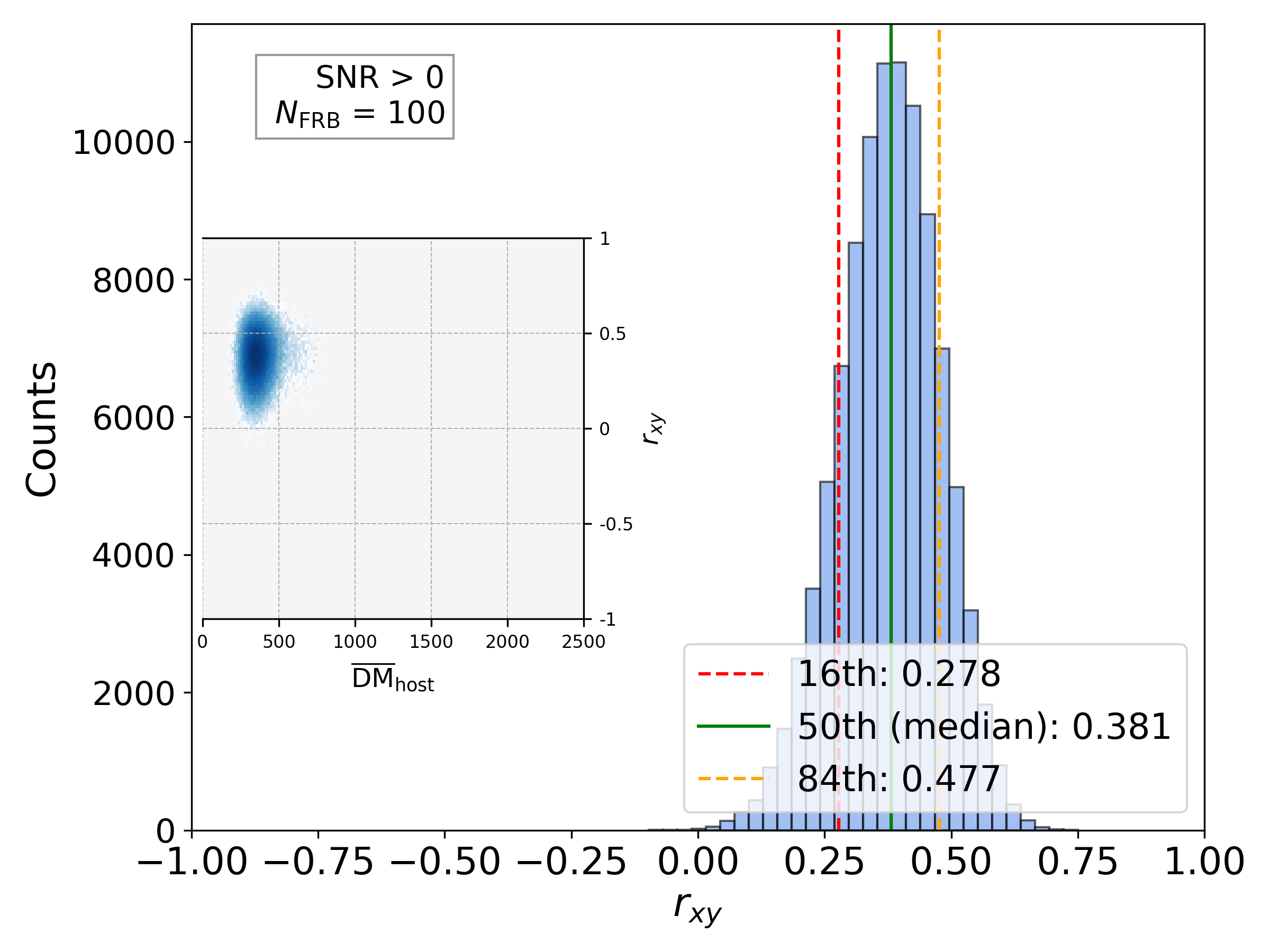}
\caption{Same as Figure~\ref{fig:corr} but considering the $\tau - {\rm DM}^2$ relation arising 
from the cloudlet model by \cite{Cordes2022} and a uniform distribution for their geometric 
and turbulent parameters.}
\label{fig:corrc22}
\end{figure*}

\begin{figure*}\centering
\includegraphics[width=0.5\textwidth]{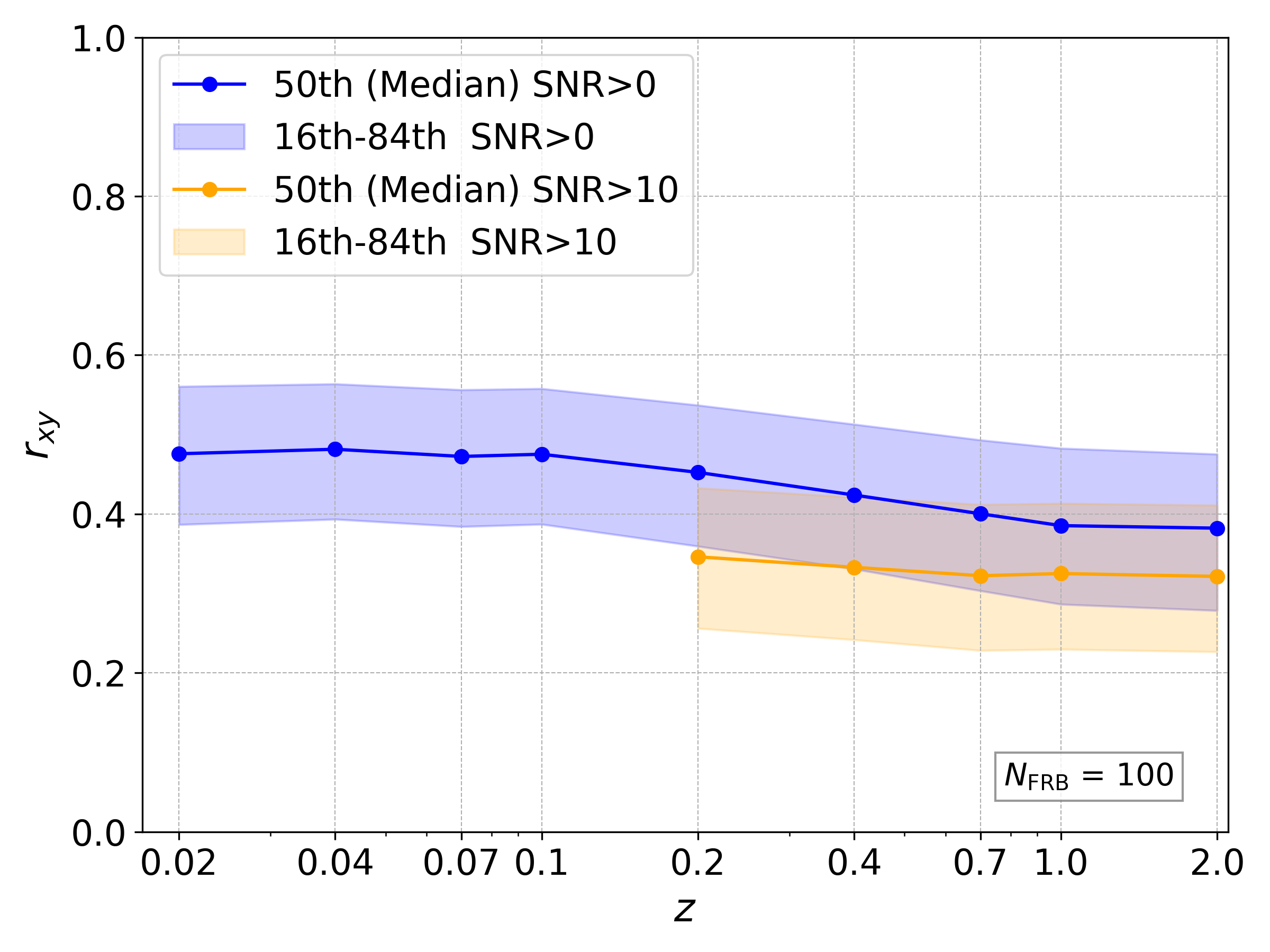}\includegraphics[width=0.5\textwidth]{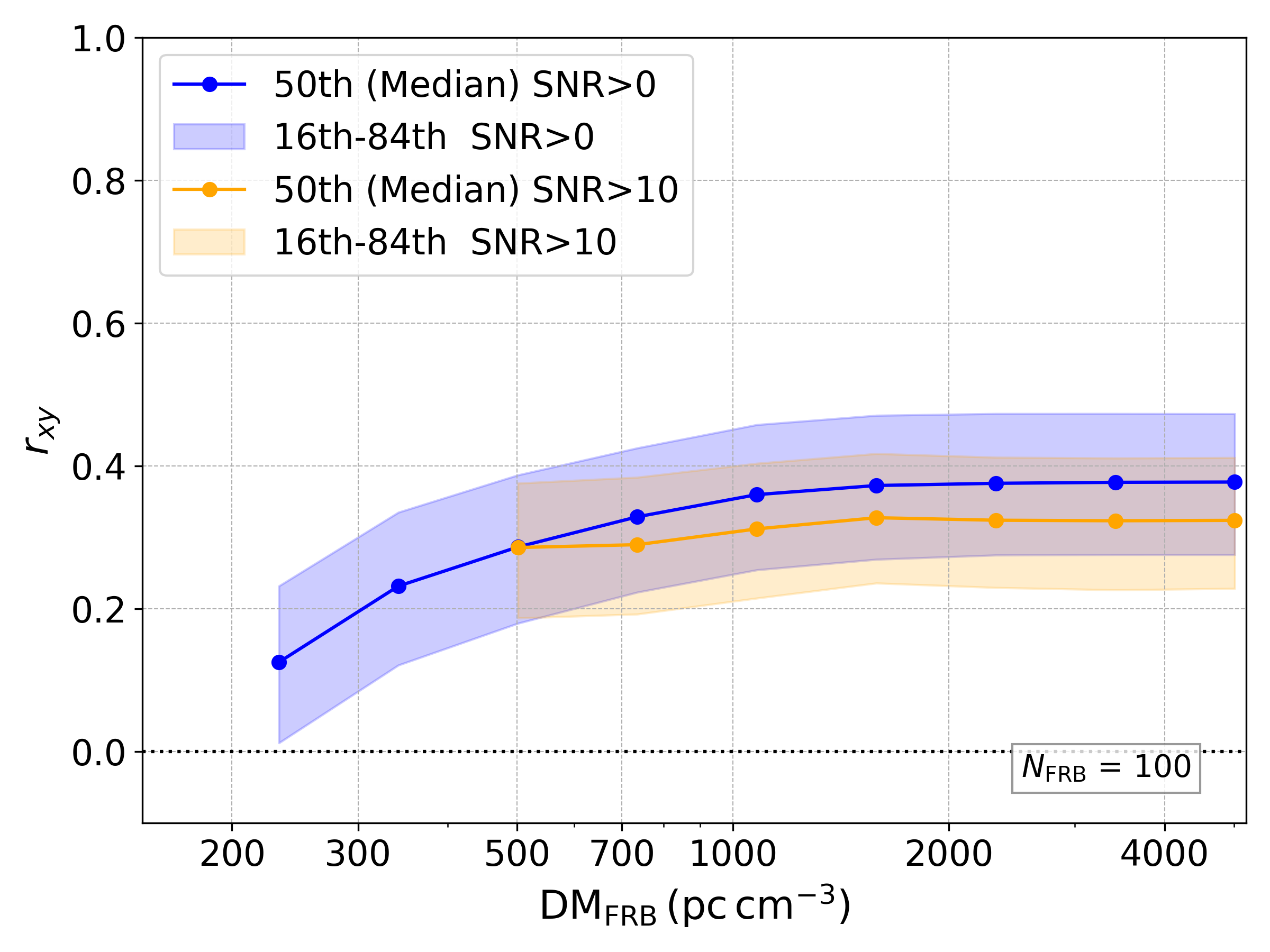}
\caption{Same as Figure~\ref{fig:zdm} but considering the $\tau - {\rm DM}^2$ relation arising 
from the cloudlet model by \cite{Cordes2022} and a uniform distribution for their geometric 
and turbulent parameters.}
\label{fig:zdmc22}
\end{figure*}

\clearpage
\bibliography{sample7}{}
\bibliographystyle{aasjournalv7}

\end{document}